\documentclass[twocolumn,showpacs,superscriptaddress,preprintnumbers,prd]{revtex4}
\usepackage{graphicx,amsmath,dcolumn,bm,color,ulem,times}
\newcommand{\Psfig}[2]{\includegraphics[width=#1]{#2}}

\renewcommand{\sout}{\bgroup \color[rgb]{1,0,0} \ULdepth=-.5ex \ULset}

\def\gev{\text{ GeV}}

\usepackage{color}

\begin{document}
\preprint{KEK-TH-1881, J-PARC-TH-0047}

\title{Constituent-counting rule in photoproduction of hyperon resonances}

\author{Wen-Chen~Chang} 
\affiliation{Institute of Physics, Academia Sinica, Taipei 11529, Taiwan}

\author{S.~Kumano}
\affiliation{KEK Theory Center,
             Institute of Particle and Nuclear Studies, \\
             High Energy Accelerator Research Organization (KEK), \\
             1-1, Ooho, Tsukuba, Ibaraki, 305-0801, Japan}
\affiliation{J-PARC Branch, KEK Theory Center,
             Institute of Particle and Nuclear Studies, KEK, \\
           and
           Theory Group, Particle and Nuclear Physics Division, 
           J-PARC Center, \\
           203-1, Shirakata, Tokai, Ibaraki, 319-1106, Japan}

\author{Takayasu~Sekihara} 
\affiliation{Research Center for Nuclear Physics
  (RCNP), Osaka University, Ibaraki, Osaka 567-0047, Japan}

\date{\today}
\begin{abstract}
  We analyze the CLAS data on the photoproduction of hyperon
  resonances, as well as the available data for the ground state
  $\Lambda$ and $\Sigma ^{0}$ of the CLAS and SLAC-E84 collaborations,
  by considering constituent-counting rule suggested by perturbative
  QCD.  The counting rule emerges as a scaling behavior of cross
  sections in hard exclusive reactions with large scattering angles,
  and it enables us to determine the number of elementary constituents
  inside hadrons. Therefore, it could be used as a new method for
  identifying internal constituents of exotic-hadron candidates. From
  the analyses of the $\gamma \, p \to K^{+} \Lambda$ and $K^{+}
  \Sigma ^{0}$ reactions, we find that the number of the elementary
  constituents is consistent with $n_{\gamma} = 1$, $n_{p} = 3$,
  $n_{K^{+}} = 2$, and $n_{\Lambda} = n_{\Sigma ^{0}} = 3$. Then, the
  analysis is made for the photoproductions of the hyperon resonances
  $\Lambda (1405)$, $\Sigma (1385)^{0}$, and $\Lambda (1520)$, where
  $\Lambda (1405)$ is considered to be a $\bar K N$ molecule and hence
  its constituent number could be five.  However, we find that the
  current data are not enough to conclude the numbers of their
  constituent. It is necessary to investigate the higher-energy region
  at $\sqrt{s} > 2.8$ GeV experimentally beyond the energy of the
  available CLAS data for counting the number of constituents clearly.
  We also mention that our results indicate energy dependence in the
  constituent number, especially for $\Lambda (1405)$.  If an excited
  hyperon is a mixture of three-quark and five-quark states, the
  energy dependence of the scaling behavior could be valuable for
  finding its composition and mixture.
\end{abstract}
\maketitle

\section{Introduction}
\label{intro}

Almost all the hadrons are described by assuming that they are
composed of minimal numbers of quarks to construct color singlet
states, i.e., three quarks ($q q q$) for baryons and a
quark--antiquark pair ($q \bar{q}$) for mesons, in conventional quark
models~\cite{Agashe:2014kda}.  However, the fundamental theory of
strong interaction, quantum chromodynamics (QCD), allows us to
consider hadrons whose structure cannot be classified into $q q q$ nor
$q \bar{q}$, such as pentaquark systems ($q q q q \bar{q}$), as long
as they are color singlet states.  They are called ``exotic'' hadrons.
One of the exotic hadron candidates is the lowest hyperon resonance,
$\Lambda (1405)$, which has been expected to be a $\bar{K} N$ bound
state rather than a $u d s$ state for a long time~\cite{Lambda-1405}.
In addition, it is very encouraging that recent experimental analyses
in heavy-quark sector discovered charged quarkonium-like
states~\cite{Belle:2011aa} and charmonium--pentaquark
states~\cite{Aaij:2015tga} as candidates of the exotic hadrons.

In order to conclude the structure of an exotic hadron candidate, it
is essential to clarify quark--gluon composition of the hadron.  For
example, the total number of constituent quarks and antiquarks is $n (
q ) + n (\bar{q} ) = 3$ for a usual $q q q$ baryon, but it becomes $n
( q ) + n ( \bar{q} ) = 5$ for a pentaquark $q q q q \bar{q}$ state.
Although one cannot, in general, count the total number of quarks and
antiquarks inside a hadron due to the annihilation and creation of a
quark--antiquark pair, there is an exception in hard exclusive
reactions with large scattering angles. There, the so-called
constituent-counting rule emerges as a scaling behavior of the cross
sections according to perturbative QCD~\cite{Brodsky:1973kr,
  Brodsky:1974vy}.  Actually, in the constituent-counting rule, the
scaling factor for the cross section corresponds to the total number
of elementary constituents involved in the hard-exclusive reaction.
This enables us to determine the number of elementary constituents
inside hadrons.

In the experimental side, the constituent-counting rule was first
applied in Ref.~\cite{Anderson:1976ph} to the reactions $\gamma \, p
\to \pi ^{+}n$, $\pi ^{0} p $, $\pi ^{-} \Delta ^{++}$, $\rho ^{0} p$,
$K^{+} \Lambda$, and $K^{+} \Sigma ^{0}$ by using the photoproduction
data of the SLAC (SLAC National Accelerator Laboratory)-E84
experiment, implying the number of the constituent $n = 3$ for the
baryons and $n = 2$ for the mesons.  The $\gamma \, p \to \pi ^{+} n$
reaction was analyzed recently by the JLab (Thomas Jefferson National
Accelerator Facility) Hall-A and E94-104 collaborations
\cite{Zhu:2002su, Zhu:2004dy}, and they found that the experimental
data are consistent with $n_{\gamma}=1$, $n_{p} = n_{n} = 3$, and
$n_{\pi ^{+}}=2$.  Recent data on the $\Lambda$ photoproduction from
CLAS at JLab were analyzed in Ref.~\cite{Schumacher:2010qx}, and the
expected $n = 9$ scaling appears to be valid for $\gamma \, p \to K^+
\Lambda$.  Various meson photoproductions from CLAS were investigated
in Ref.~\cite{Dey:2014mka}, but the author could not conclude the
scaling for the vector mesons as suggested in conventional quark
models due to additional gluon exchanges from higher Fock states for
hadrons.  The constituent-counting rule was also confirmed by two-body
hadronic exclusive reactions at BNL (Brookhaven National
Laboratory)-AGS (Alternating Gradient
Synchrotron)~\cite{bnl-exclusive}.  It is noteworthy that the
constituent-counting rule was experimentally shown to be valid even
for nuclei~\cite{Uzikov:2005jd, Ilieva:2013xya}; the numbers of
elementary constituents inside the deuteron, ${}^{3}\text{H}$, and
${}^{3}\text{He}$ are counted as $n_{d} = 6$ and $n_{{}^{3}\text{H}} =
n_{{}^{3}\text{He}} = 9$.

On the other hand, there are theoretical proposals recently to use the
constituent-counting rule for clarifying the internal structure of
hadrons, especially exotic-hadron candidates~\cite{Kawamura:2013iia,
  Kawamura:2013wfa, Galynskii:2013esa, Blitz:2015nra,
  Brodsky:2015wza}.  Actually, if a baryon (meson) contains more than
three (two) constituents as an exotic hadron, it should explicitly
affect the scaling behavior of its production cross section at high
energies.  Moreover, since the number of elementary constituents
inside nuclei is counted as $n_{d} = 6$ and $n_{{}^{3}\text{H}} =
n_{{}^{3}\text{He}} = 9$~\cite{Uzikov:2005jd, Ilieva:2013xya}, we can
apply the constituent-counting rule not only to compact exotic hadrons
but also to hadronic molecules composed of two (or more) color singlet
states, although we cannot distinguish whether the hadron is a compact
multiquark system or a hadronic molecule solely by the
constituent-counting rule.  For instance, the number of constituents
inside $\Lambda (1405)$ is considered to be five if it is a $\bar{K}
N$ bound state, which cannot be distinguished from a compact
pentaquark state only by the scaling behavior.  Nevertheless, there
are possibilities that compact multiquark and diffuse hadron-molecule
states could be further distinguished by the so-called
compositeness~\cite{Hyodo:2011qc, Hyodo:2013nka, Sekihara:2013sma,
  Sekihara:2014kya} and hadron tomography in terms of
three-dimensional structure functions, generalized parton
distributions (GPDs), transverse-momentum-dependent parton
distributions (TMDs), and generalized distribution amplitudes (GDAs)
\cite{Kawamura:2013wfa}.  However, much efforts should be made for
such tomography studies because there is no experimental information
on these observables for exotic hadron candidates at this stage.

In this article, we use the constituent-counting rule as a guiding
principle in analyzing available experimental data on hard exclusive
cross sections at high energies. Particularly, we investigate new data
obtained by the CLAS collaboration at JLab for the photoproduction of
the hyperon resonances $\Lambda (1405)$, $\Sigma (1385)^{0}$, and
$\Lambda (1520)$ from the proton target ($\gamma \, p \to K^{+} Y^{(
  \ast )}$) \cite{Moriya:2013hwg}, extending the preceding studies on
the scaling behavior for the ground-state hyperons.  Among the hyperon
resonances, $\Lambda (1405)$ is expected to be a $\bar K N$
molecule~\cite{Lambda-1405} and hence its constituent number could be
five, so that it is interesting to obtain the constituent numbers of
these exotic hadron candidates.  The hyperons $\Sigma (1385)^{0}$ and
$\Lambda (1520)$ are commonly considered as ordinary baryons with the
$qqq$ configuration; however, there are some theoretical studies that
$\Lambda (1520)$ may have a certain fraction of hadronic molecular
components~\cite{Aceti:2014wka}.  We also analyze new CLAS data for
$\Lambda$ and $\Sigma^0$~\cite{McCracken:2009ra,Dey:2010hh} and the
SLAC-E84 data~\cite{Anderson:1976ph}.  In the CLAS experiments, the
photon momentum in the laboratory frame is up to $3.8 \gev / c$ and
hence the center-of-mass energy is up to $2.8 \gev$.  We check the
scaling behavior of the photoproduction cross sections and investigate
a possibility to count the elementary constituents inside the hadrons.

This article is organized in the following way.  In
Sec.~\ref{counting}, the constituent-counting rule and the cross
section is briefly explained.  The photoproduction cross-section data
are analyzed, and cross-section slopes $1/s^{n-2}$ are examined at
high energies to show the constituent numbers $n$ in
Sec.~\ref{results}.  We summarize our studies in Sec.~\ref{summary}.

\section{Constituent-counting rule in hard exclusive reactions}
\label{counting}

We introduce the constituent counting rule for a hard exclusive
reaction by showing expressions for its cross section and matrix
element and explaining a typical hard-gluon-exchange process
which contributes to the exclusive reaction.  The cross section for an
exclusive reaction $a+b \to c+d$ is given by the matrix element $M_{ab
  \to cd}$ as
\begin{align}
\frac{d\sigma_{ab \to cd}}{dt}
& \simeq \frac{1}{16 \pi s^2}
\overline{\sum_{\rm pol}} \, | M_{ab \to cd} |^2 .
\label{eqn:two-body-cross}
\end{align}
Here, the approximation means that the masses of particles are
neglected, and the Mandelstam variables $s$ and $t$ are defined by the
momenta of the particles, $p_i$ ($i=a,\,b,\,c,\,d$), as
\begin{align}
& 
s = (p_a + p_b)^2 
    \simeq 4 \, | \, \vec{p}_{\rm cm} \, |^2, 
\nonumber \\
& 
t = (p_a - p_c)^2 
    \simeq -2 \, | \, \vec{p}_{\rm cm} \, |^2 (1-\cos \theta_{\rm cm}), 
\label{eqn:st}
\end{align}
where $p_{\rm cm}$ and $\theta_{\rm cm}$ are momentum and scattering
angle in the c.m.\ frame, respectively.  The average over the initial
spins and the summation for the final spins are taken in
Eq.~\eqref{eqn:two-body-cross}.

\begin{figure}[b!]
  \centering
  \Psfig{5.5cm}{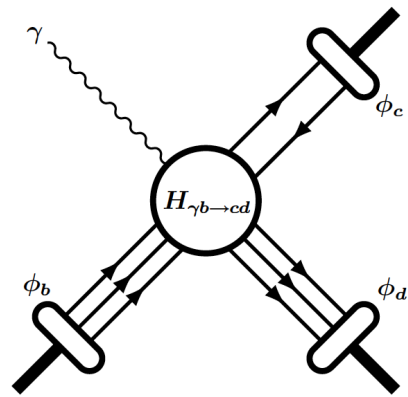}
\vspace{-0.2cm}
  \caption{Hard exclusive reaction $\gamma + b \to c +d$.}
\label{fig:exclusive-abcd}
\end{figure}

In this article, we investigate hard photoproductions of
hadrons, so that the reaction $\gamma + b \to c +d$ is considered
in the following discussions.
The matrix element is denoted in general by
the photon-parton scattering amplitude $H_{\gamma b \to cd}$
and the light-cone distribution amplitudes of hadrons,
$\phi_b$, $\phi_c$, and $\phi_d$, as
\cite{Mueller-Brodsky}
\begin{align} 
M_{\gamma b \to cd} = & \int [dx_b] \, [dx_c] \, [dx_d]  \,
    \phi_c ([x_c]) \, \phi_d ([x_d]) 
\nonumber \\
& 
\times 
H_{\gamma b \to cd} ([x_b],[x_c],[x_d],Q^2) \, \phi_b ([x_b]) ,
\label{eqn:mab-cd}
\end{align}
as shown in Fig.~\ref{fig:exclusive-abcd}.  Here, $[x]$ indicates a
set of the light-cone momentum fractions of partons in a hadron,
$x_j=p_j^+/p^+$ with $j$th parton and hadron momenta $p_j$ and $p$,
respectively.  The light-cone component is defined as $p^+ = (p^0 +
p^3)/\sqrt{2}$ with the choice of the third axis for the longitudinal
direction.  The hadron distribution amplitude $\phi_i ([x])$ is the
amplitude for finding quarks with the momentum fractions $[x]$ in the
hadron $i$.

\begin{figure}[b!]
\vspace{-0.3cm}
  \centering
  \Psfig{7.5cm}{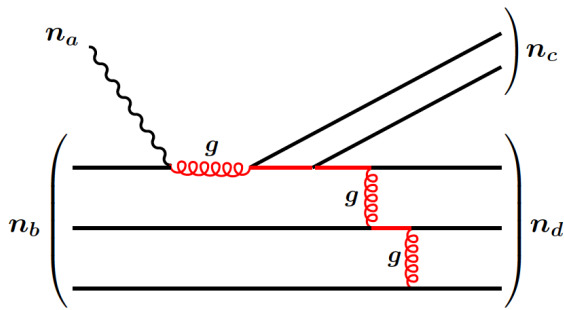}
\vspace{-0.2cm}
  \caption{(color online) A typical hard-gluon-exchange process
   in the exclusive reaction $\gamma + b \to c +d$
  at high energies.}
\label{fig:exclusive-proc}
\end{figure}

A hard exclusive process $\gamma + b \to c + d$ can occur if a large
momentum transfer is shared by exchanged hard gluons as shown
in Fig.~\ref{fig:exclusive-proc}. Considering hard gluon 
and quark propagators in the intermediate state together
with normalization factors of initial and final quarks,
we have the scaling property for its cross section at high-momentum
transfer in terms of the number of constituents which participate
in the reaction \cite{Kawamura:2013iia}.
Its cross section is expressed
as~\cite{Brodsky:1973kr, Brodsky:1974vy}
\begin{equation}
\frac{d \sigma}{d t} = \frac{1}{s^{n - 2}} \, f ( t / s ) ,
\label{eq:scale}
\end{equation}
by the Mandelstam variables $s$ and $t$.  Here, $f(t / s)$ is a
function of a variable $t/s$ which corresponds to the center-of-mass
scattering angle $\theta _{\rm cm}$ in the high-energy region [see
  Eq.~\eqref{eqn:st}].  The scaling factor $n$ counts the number of
the participating constituents. In the $\gamma + b \to c + d$
reaction, we have $n= n_{\gamma} + n_{b} + n_{c} + n_{d}$, where the
particle $i$ ($= \gamma$, $b$, $c$, and $d$) is made of $n_{i}$
elementary constituents. Therefore, from the scaling law of the cross
section~\eqref{eq:scale}, we can learn the total number of constituents
$n$. The detailed formulation of the constituent-counting rule is
found in Ref.~\cite{Kawamura:2013iia}.

Although the constituent-counting rule is a useful guideline, we do
not know the energy at which the scaling starts.  If the absolute
cross section were calculated precisely from the low-energy resonance
region to the high-energy perturbative QCD region, the transition
point should be determined.  The resonance region could be
theoretically described relatively well by effective hadron
models. However, it is not the case in the high-energy region,
although the high-energy slope should be determined by the constituent
counting rule in Eq.~\eqref{eq:scale}.

There are two major difficulties in calculating the absolute cross
section in Eq.~\eqref{eqn:two-body-cross}. The hadron distribution
amplitudes $\phi_i ([x])$ are not well investigated except for the
pion \cite{B-factory}, so that the amplitude $H_{\gamma b \to cd}
([x_b],[x_c],[x_d],Q^2)$ cannot be calculated reliably. Second, there
are many diagrams which contribute to the exclusive reaction depending
how the gluons couple to quarks.  For instance, we show a typical
diagram for the process $\gamma \, p \to \pi^+ n$ in
Fig.~\ref{fig:exclusive-proc}, which does not look complicated.
However, as the quark numbers grow especially in exotic hadrons, the
number of diagrams increases significantly, so that development of an
automatic code to obtain Feynman diagrams and to estimate their
numerical contributions is essential. Unless we complete these serious
studies, it is theoretically impossible to calculate the absolute
magnitude of an exclusive cross section, hence to obtain the minimum
energy for the scaling.

In spite of these issues for estimating the absolute cross section,
the constituent counting rule itself is considered to be valid by
perturbative QCD for a hard exclusive reaction.  We use this scaling
rule for finding the numbers of elementary constituents which
participate in the reaction. This should be used for exotic hadron
studies because the number of constituent quarks should be different
from ordinary hadrons ($q\bar q$, $qqq$).

\section{Analysis results}
\label{results}

In this study we consider the $\gamma \, p \to K^{+} Y^{( \ast )}$
reactions with $Y^{( \ast )}$ being the $\Lambda$, $\Sigma ^{0}$,
$\Lambda (1405)$, $\Sigma (1385)^{0}$, and $\Lambda (1520)$ hyperons.
The reactions $\gamma \, p \to K^{+} \Lambda$ and $K^{+} \Sigma ^{0}$ are
observed in SLAC-E84~\cite{Anderson:1976ph} and
CLAS~\cite{McCracken:2009ra, Dey:2010hh}, and the reactions $\gamma \, p
\to K^{+} \Lambda (1405)$, $K^{+} \Sigma (1385)^{0}$, and $K^{+}
\Lambda (1520)$ in CLAS~\cite{Moriya:2013hwg}.  In the CLAS
experiment, they measured the differential cross sections $d \sigma /
d \cos \theta _{\rm cm}$.  This can be translated into the
differential cross section $d \sigma /d t$ by the formula
\begin{equation}
\frac{d \sigma}{d t} = \frac{1}{2 \, p_{\rm in} \, p_{\rm out}} 
\frac{d \sigma}{d \cos \theta _{\rm cm}} , 
\end{equation}
where $p_{\rm in}$ ($p_{\rm out}$) is the center-of-mass momenta of
the particles in the initial (final) state. Since the scaling behavior
appears in the large scattering-angle region, we take the data in
scattering-angle bins around the right angle $\theta _{\rm cm} =
90^{\circ}$ in the following analyses.  Here we note that, as
explained in the previous section, we do not know beforehand the
minimum energy for the scaling, so that in the present analyses we
examine several values of the minimal energy for the scaling,
$\sqrt{s_{\rm min}}$, on a case-by-case basis for each reaction.  We
did not analyze the $\gamma \, n \to K^{+} \Sigma ^{-}$ cross section
from CLAS~\cite{AnefalosPereira:2009zw} because for its measurement
they used bound neutron in the deuteron, for which we may need to take
into account nuclear effects.

\begin{figure}[t]
  \centering
  \Psfig{8.6cm}{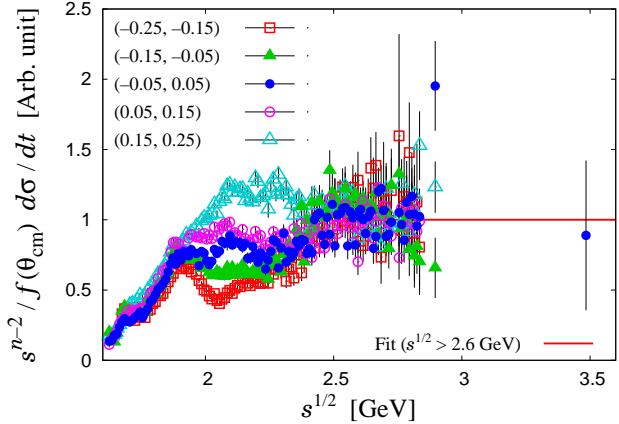}
\vspace{-0.5cm}
\caption{(color online) Experimental data of $\gamma \, p \to K^{+}
  \Lambda$ cross section $d \sigma / d t$~\cite{McCracken:2009ra,
    Anderson:1976ph} multiplied by $s^{n - 2} / f ( \theta _{\rm cm})$
  ($n=10.0$) with the function $f ( \theta _{\rm cm} )$ in
  Eq.~\eqref{eqn:f-theta}.  The scaling factor $n$ is fixed at $10.0$
  by fitting in the region $\sqrt{s} \ge 2.6 \gev$ as shown in
  Table~\ref{tab:n1}.  The solid line indicates a fit to the data. }
\label{fig:KpL}
\end{figure}

\begin{figure}[t!]
  \centering
  \Psfig{8.6cm}{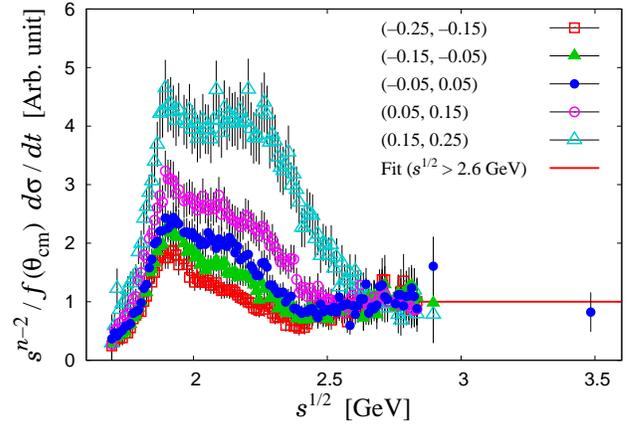}
\vspace{-0.5cm}
\caption{(color online) Experimental data of $\gamma \, p \to K^{+}
  \Sigma ^{0}$ cross section $d \sigma / d t$~\cite{Dey:2010hh,
    Anderson:1976ph} multiplied by $s^{n - 2} / f ( \theta _{\rm cm})$
  ($n=9.2$) with the function $f ( \theta _{\rm cm} )$ in
  Eq.~\eqref{eqn:f-theta}.  The scaling factor $n$ is fixed at $9.2$
  by fitting in the region $\sqrt{s} \ge 2.6 \gev$ as shown in
  Table~\ref{tab:n1}.  The solid line is a fit to the data. }
\label{fig:KpS}
\end{figure}

\begin{figure}[!t]
  \centering
  \Psfig{8.6cm}{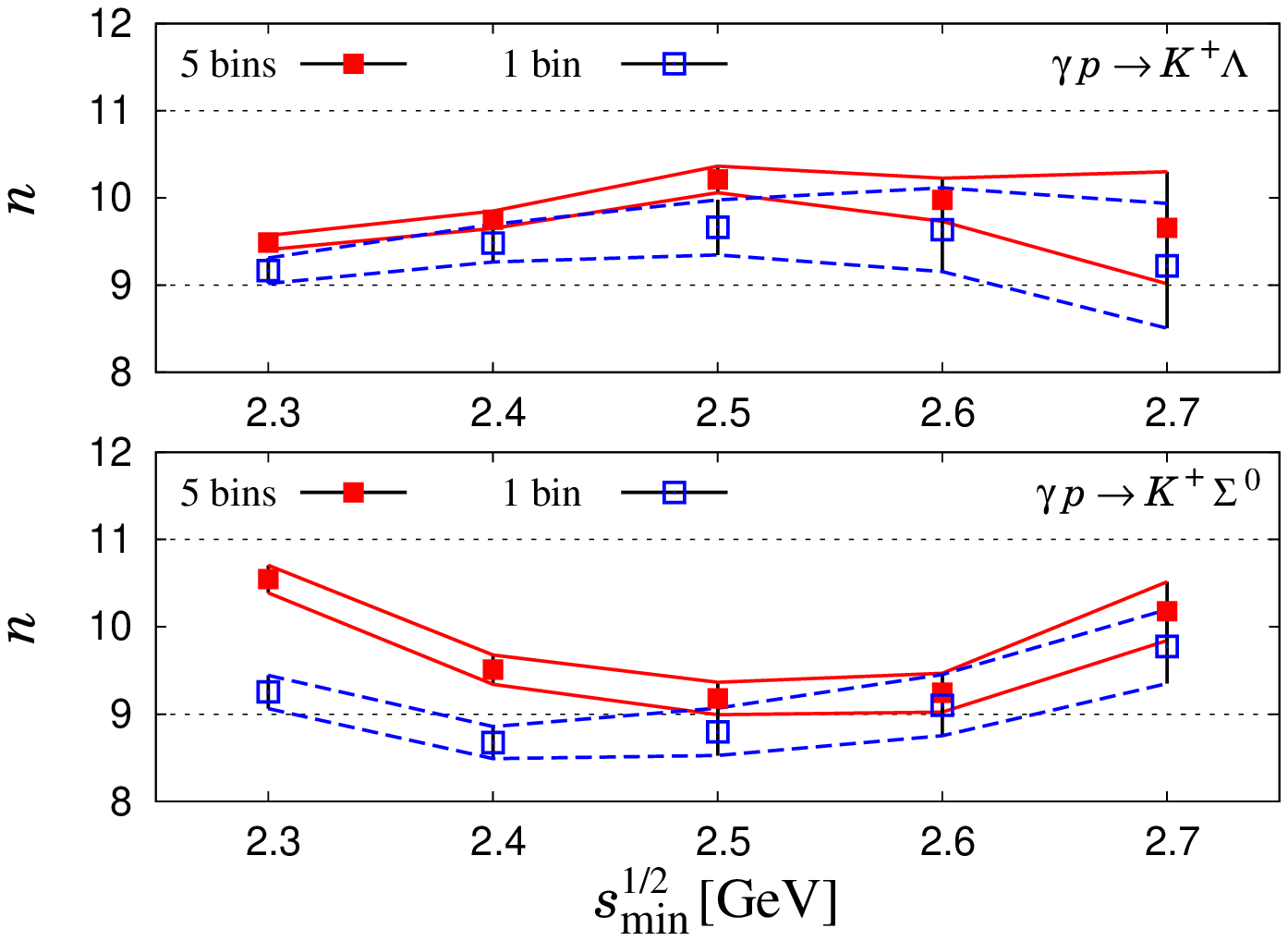}
\vspace{-0.8cm}
  \caption{(color online) Behavior of the scaling factor $n$ by
    changing the minimal value of the energy $\sqrt{s_{\rm min}}$
    for fitting. }
\label{fig:n_LS}
\end{figure}

\begin{figure}[t]
  \centering
  \Psfig{8.6cm}{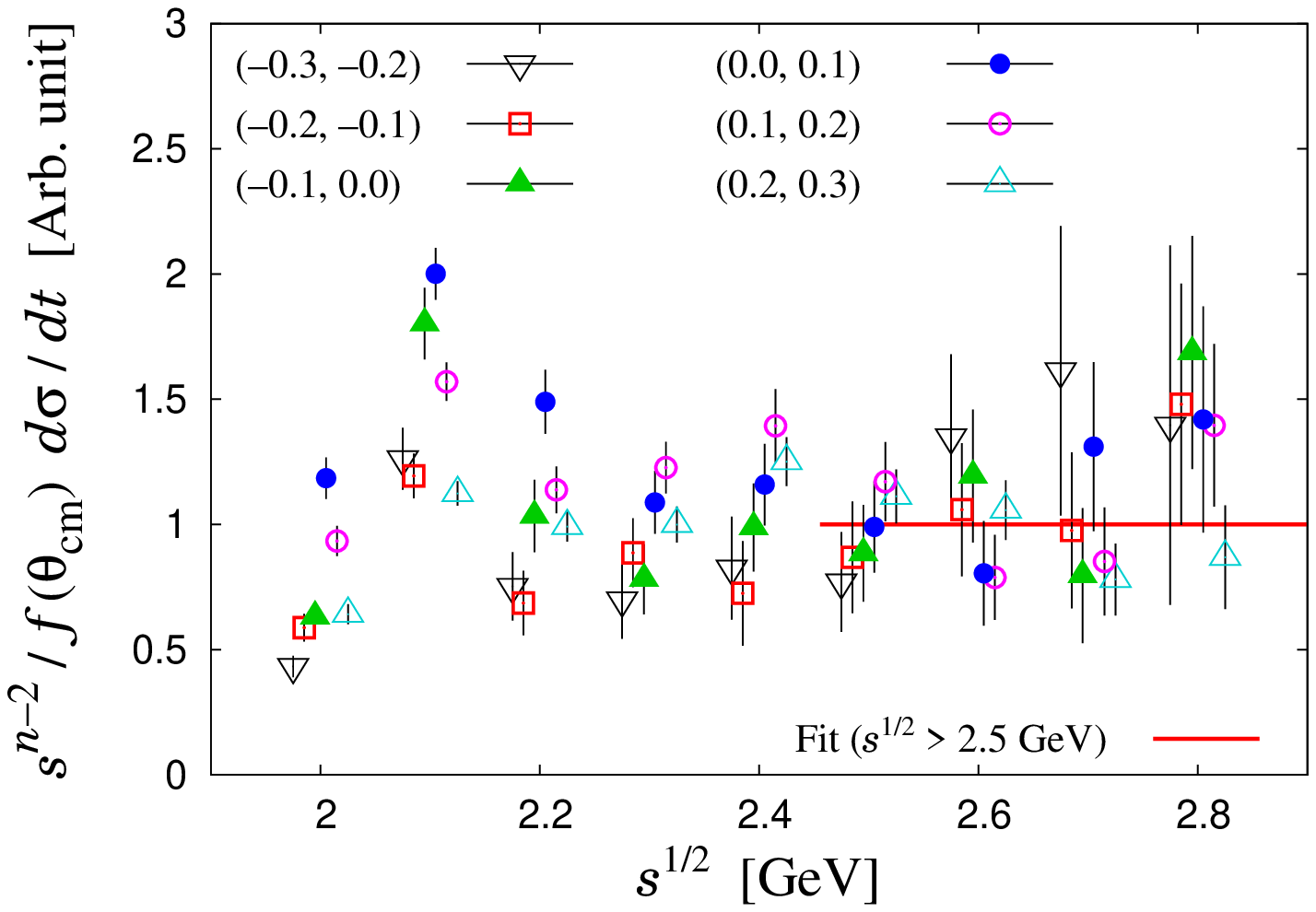}
\vspace{-0.5cm}
\caption{(color online) Experimental data of $\gamma \, p \to K^{+}
  \Lambda (1405)$ cross section $d \sigma / d t$~\cite{Moriya:2013hwg}
  multiplied by $s^{n - 2} / f ( \theta_{\rm cm} )$ ($n=10.6$) with
  the function $f ( \theta _{\rm cm} )$ in Eq.~\eqref{eqn:f-theta}.
  The scaling factor $n$ is fixed at $10.6$ by fitting in the region
  $\sqrt{s} \ge 2.5 \gev$ as shown in Table~\ref{tab:n2}.  The solid
  line is a fit to the data.  For a better visualization, we slightly
  shift the data to horizontal direction.}
\label{fig:KpL1405}
\end{figure}

\begin{figure}[t!]
\vspace{-0.05cm}
  \centering
  \Psfig{8.6cm}{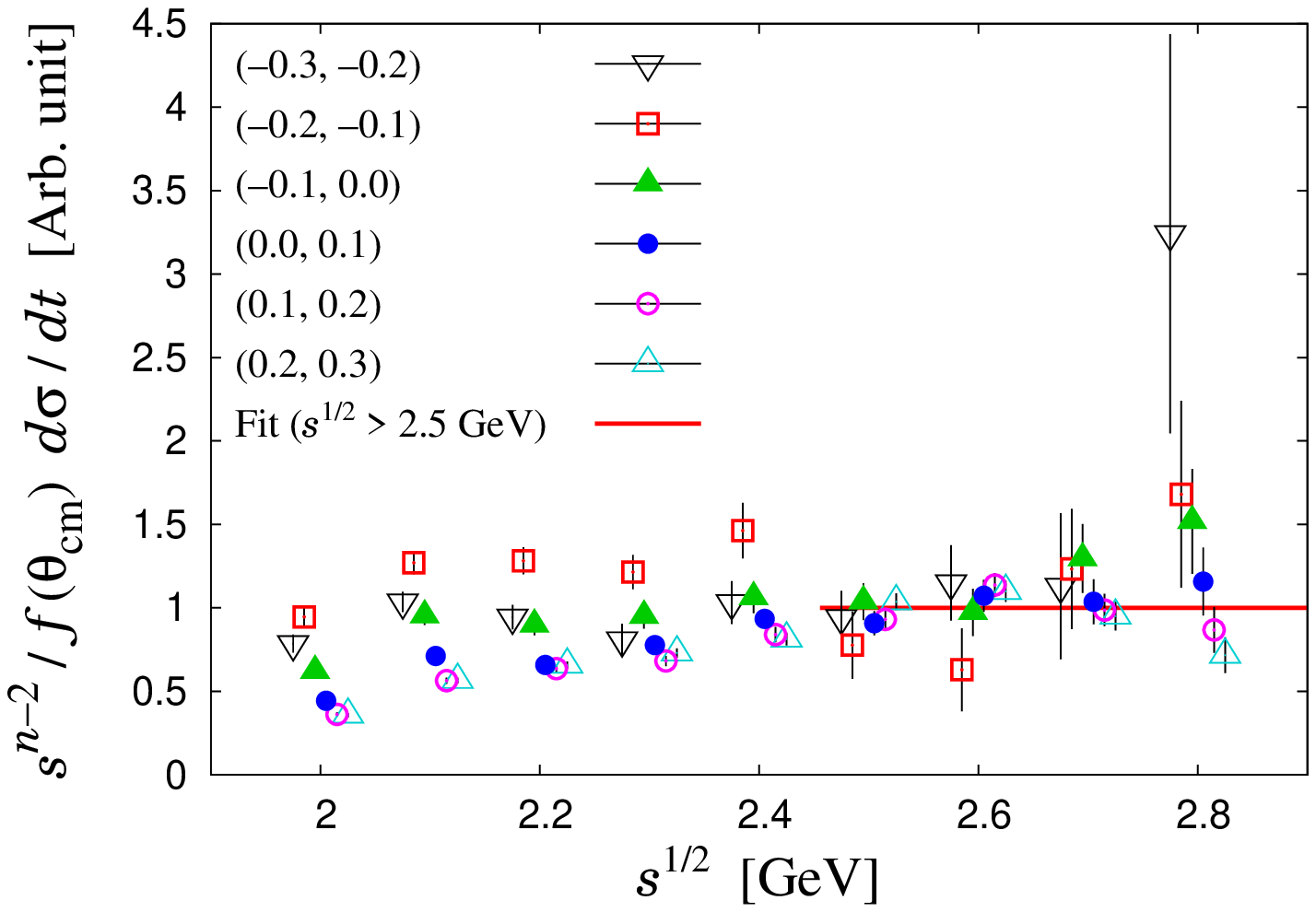}
\vspace{-0.5cm}
\caption{(color online) Experimental data of $\gamma \, p \to K^{+}
  \Sigma (1385)^{0}$ cross section $d \sigma / d
  t$~\cite{Moriya:2013hwg} multiplied by $s^{n - 2} / f ( \theta_{\rm
    cm} )$ ($n=11.4$) with the function $f ( \theta _{\rm cm} )$ in
  Eq.~\eqref{eqn:f-theta}.  The scaling factor $n$ is fixed at $11.4$
  by fitting in the region $\sqrt{s} \ge 2.5 \gev$ as shown in
  Table~\ref{tab:n2}.  The solid line is a fit to the data.  For a
  better visualization, we slightly shift the data to horizontal
  direction.}
\label{fig:KpS1385}
\end{figure}

\begin{figure}[t!]
\vspace{-0.05cm}
  \centering
  \Psfig{8.6cm}{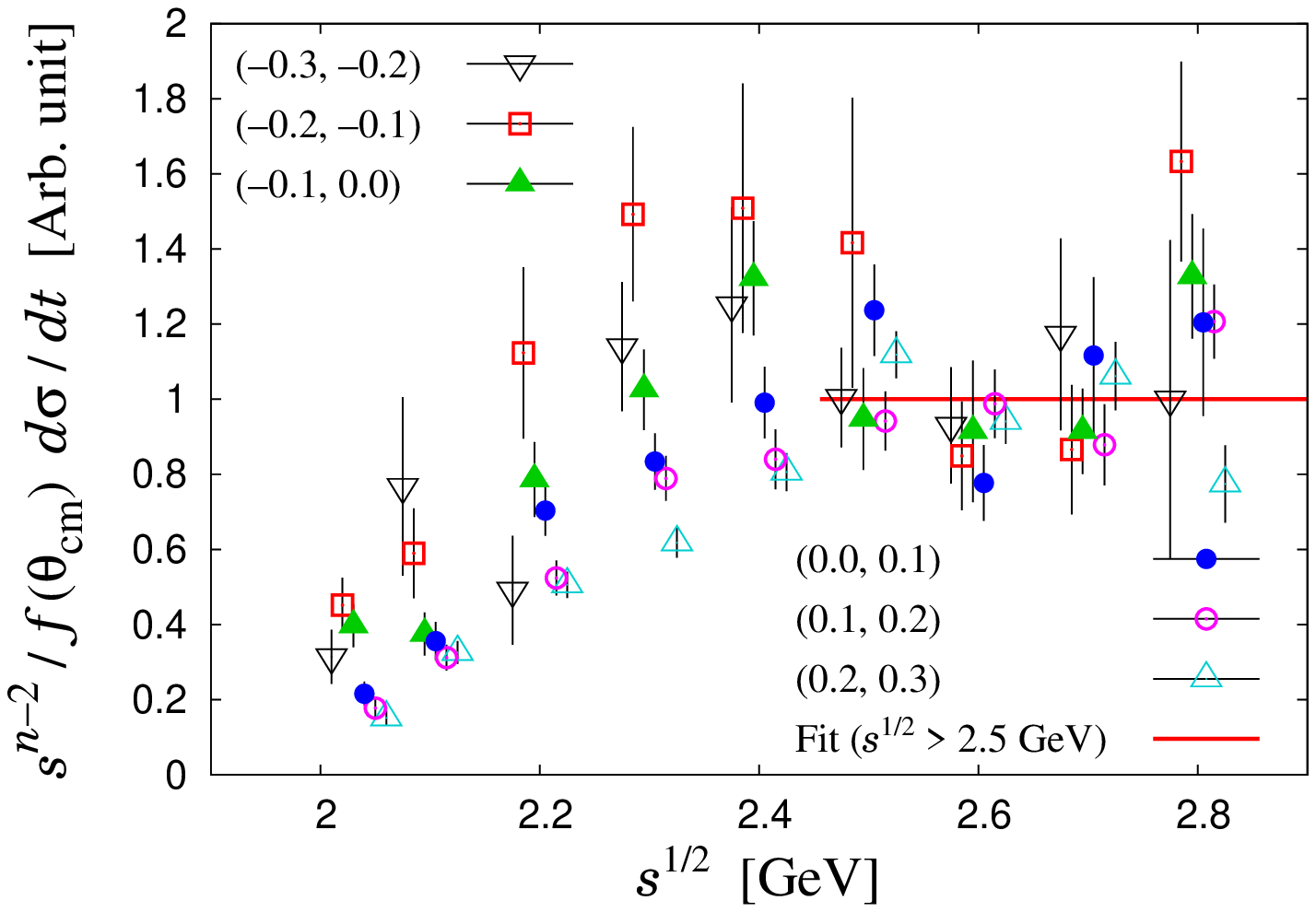}
\vspace{-0.5cm}
\caption{(color online) Experimental data of $\gamma \, p \to K^{+}
  \Lambda (1520)$ cross section $d \sigma / d t$~\cite{Moriya:2013hwg}
  multiplied by $s^{n - 2} / f ( \theta_{\rm cm} )$ ($n=9.8$) with the
  function $f ( \theta _{\rm cm} )$ in Eq.~\eqref{eqn:f-theta}.  The
  scaling factor $n$ is fixed at $9.8$ by fitting in the region
  $\sqrt{s} \ge 2.5 \gev$ as shown in Table~\ref{tab:n2}.  The solid
  line is a fit to the data.  For a better visualization, we slightly
  shift the data to horizontal direction.}
\vspace{-0.00cm}
\label{fig:KpL1520}
\end{figure}

\begin{figure}[t!]
  \centering
  \Psfig{8.6cm}{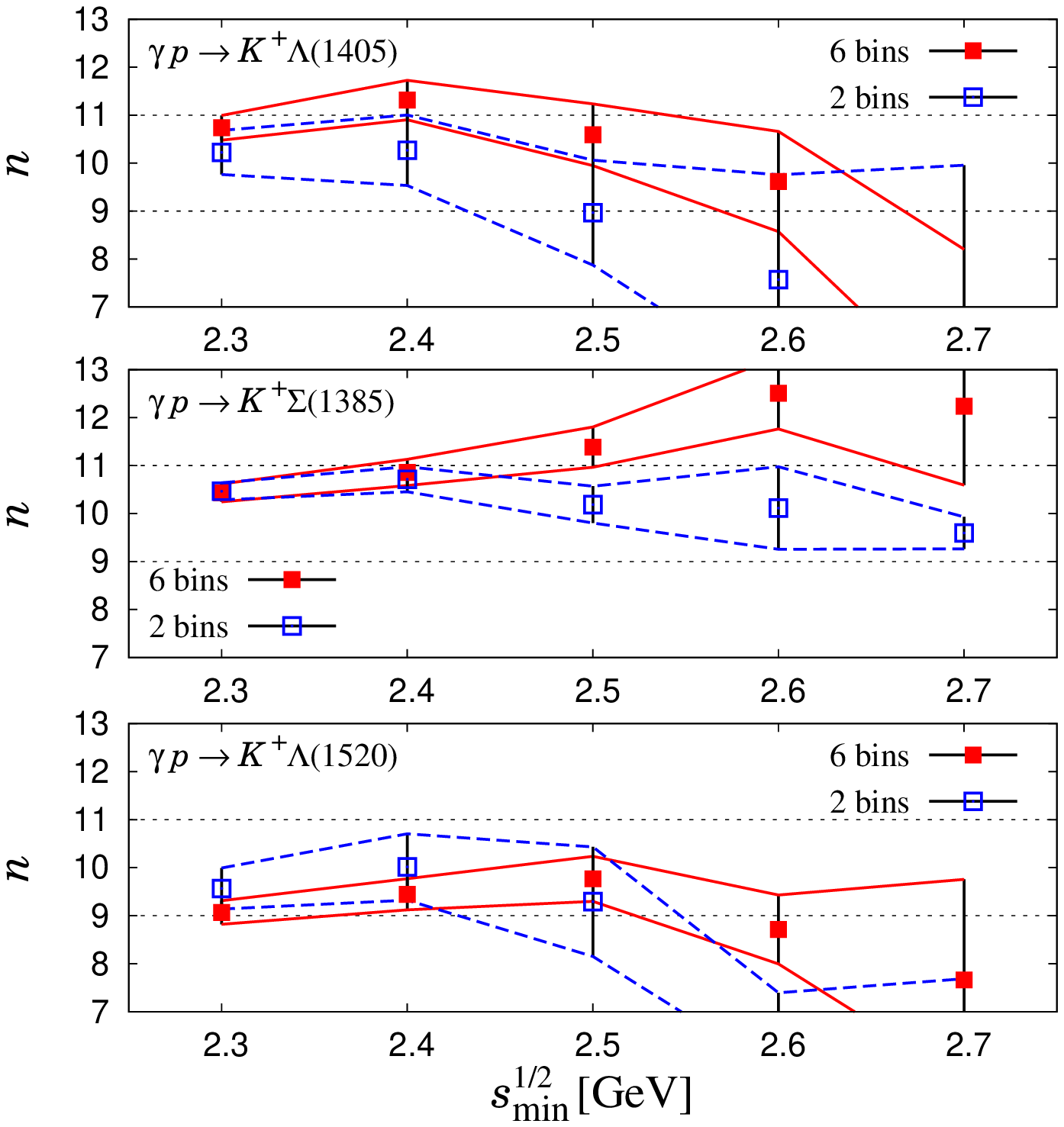}
\vspace{-0.8cm}
  \caption{(color online) Behavior of the scaling factor $n$ by
    changing the minimal value of the energy $\sqrt{s_{\rm min}}$
    for fitting. }
\label{fig:n_Ystar}
\end{figure}

The analyses are done in the following way.
We assume that the differential cross section $d \sigma /d t$ is
given by the form
$f ( \theta _{\rm cm} ) / s^{n-2}$ 
with the scattering-angle-dependent function $f(\theta _{\rm cm} )$. 
The function $f ( \theta _{\rm cm} )$ takes different values 
for different bins:
\begin{equation}
  f ( \theta _{\rm cm} )_{(1)}, \, 
  f ( \theta _{\rm cm} )_{(2)} , \, \cdots, \,
  f ( \theta _{\rm cm} )_{(N)}  ,
\label{eqn:f-theta}
\end{equation}
in fitting $N$ bins of the scattering angle.
Here, the constants 
$f ( \theta _{\rm cm} )_{(1)}$,  $\cdots$,
$f ( \theta _{\rm cm} )_{(N)}$
are also fitting parameters.  
The scaling factor $n$ is fixed regardless of the scattering angle. 
Namely, the differential cross section $d \sigma / d t$ is fitted
by the function $f ( \theta _{\rm cm} ) / s^{n-2}$ with 
$N + 1$ parameters, 
$f ( \theta _{\rm cm} )_{(1)}$,  $\cdots$,
$f ( \theta _{\rm cm} )_{(N)}$, and $n$.  
In the fitting procedure, we mean 
$5$ bins for the bins from $-0.25 < \cos \theta _{\rm cm} < -0.15$
to $0.15 < \cos \theta _{\rm cm} < 0.25$ with $0.10$ step each;
$4$ bins from $-0.2 < \cos \theta _{\rm cm} < -0.1$ 
to $0.1 < \cos \theta _{\rm cm} < 0.2$;
$2$ bins from $-0.1 < \cos \theta _{\rm cm} < 0.0$ 
to $0.0 < \cos \theta_{\rm cm} < 0.1$;
and $1$ bin for $-0.05 < \cos \theta _{\rm cm} < 0.05$.
The results are shown in Figs.~\ref{fig:KpL}, \ref{fig:KpS},
\ref{fig:n_LS}, \ref{fig:KpL1405}, \ref{fig:KpS1385},
\ref{fig:KpL1520}, and \ref{fig:n_Ystar},
and Tables~\ref{tab:n1} and \ref{tab:n2}.

First, there are many data on the photoproductions of $\Lambda$ and
$\Sigma^0$ as shown in Figs.~\ref{fig:KpL} and \ref{fig:KpS}. There
are five sets of data in each figure. For the $\Lambda$ production in
Fig.~\ref{fig:KpL}, the scaling is approximately seen for the bin of
$-0.05 < \cos \theta _{\rm cm} < 0.05$, by including the high-energy
data at 3.5 GeV from the SLAC-E84 experiment.  Other bin data deviate
slightly form the scaling behavior of $n \simeq 10$ suggested by the
$-0.05 < \cos \theta _{\rm cm} < 0.05$ bin data. Obviously, we need
measurements to fill the gap in the region, 3.0 GeV $< \sqrt{s} <$ 3.5
GeV between the existing data points.
For the $\Sigma^0$ production in Fig.~\ref{fig:KpS}, the discrepancies
between the data of different bins are more pronounced at low energies
($\sqrt{s}<2.3$ GeV).  However, they seem to merge into the scaling
function at $\sqrt{s} >$ 2.6 GeV.  The scaling is rather clear in the
data of the bin $-0.05 < \cos \theta _{\rm cm} < 0.05$ if the data at
$\sqrt{s} =$ 3.5 GeV is taken into account.  However, more data should
be obtained in the energy region $3.0 \gev < \sqrt{s} < 3.5 \gev$ for
drawing a clearer conclusion on the scaling behavior in the same way
with the $\Lambda$ production in Fig.~\ref{fig:KpL}.

Next, we determine the scaling factor $n$ from the high-energy data at
$s \ge s_{\rm min}$ by excluding low-energy resonances.  As expected
from Figs.~\ref{fig:KpL} and \ref{fig:KpS}, the obtained factor $n$
depends on the choice of $s_{\rm min}$.  By using all the five-bin
data or the one-bin data for the $\Lambda$ and $\Sigma^0$ productions,
we determine the factor $n$ and the results are shown in 
Table~\ref{tab:n1} and Fig.~\ref{fig:n_LS} for various choices of
$\sqrt{s_{\rm min}}$ from 2.3 GeV to 2.7 GeV.  If the one-bin ($-0.05
< \cos \theta _{\rm cm} < 0.05$) data are used, the determined number
is consistent with $n=9$, namely with the assignment $n_\gamma =1$,
$n_p =3$, $n_{K^+} =2$, $n_{\Lambda} =3$, and $n_{\Sigma^0} =3$
although the detailed values depend on $s_{\rm min}$.  Even if the
five-bin data are used, the determined values of $n$ are almost same
by considering errors except for the ones at $\sqrt{s} \ge
\sqrt{s_{\rm min}}=2.3$ and 2.4 GeV for $\Sigma^0$.  These results
indicate that the constituent numbers determined by the
constituent-counting rule are consistent with the standard quark model
compositions, $n_{\Lambda} =3$ and $n_{\Sigma^0} =3$.

One of the major purposes of this work is to apply the
constituent-counting rule to exotic-hadron candidates for determining
the numbers of constituents for them.  It should be a significant step
in the study of exotic hadrons to apply a new approach, which is
rather different from the current ones with low-energy observables
such as mass and decay width, to clarify their internal structure.

\begin{table}[t!]
  \caption{Determined values of the scaling factor $n$ for 
    $\gamma \, p \to K^{+} \Lambda$ and $K^{+} \Sigma ^{0}$. 
    We also show the $\chi ^{2}$ values divided by 
    the degrees of freedom $N_{\rm d.o.f.}$.
    }
  \label{tab:n1}
  \begin{ruledtabular}
    \begin{tabular*}{8.6cm}{@{\extracolsep{\fill}}lll}
      \multicolumn{3}{c}{$\gamma \, p \to K^{+} \Lambda$} \\
      $\sqrt{s_{\rm min}}$ & 
      $n$ ($\chi ^{2} / N_{\rm d.o.f.}$), 5 bins & 
      $n$ ($\chi ^{2} / N_{\rm d.o.f.}$), 1 bin
      \\
      \hline
      $2.3 \gev$ &
      \phantom{0}$9.5 \pm 0.1$ ~~ ($527 / 257$) &
      $9.2 \pm 0.1$ ~~ ($71 / 52$) 
      \\
      $2.4 \gev$ &
      \phantom{0}$9.7 \pm 0.1$ ~~ ($252 / 207$) &
      $9.5 \pm 0.2$ ~~ ($55 / 42$) 
      \\
      $2.5 \gev$ &
      $10.2 \pm 0.2$ ~~ ($151 / 157$) &
      $9.7 \pm 0.3$ ~~ ($35 / 32$) 
      \\
      $2.6 \gev$ &
      $10.0 \pm 0.2$ ~~ ($94 / 107$) &
      $9.6 \pm 0.5$ ~~ ($22 / 22$)
      \\
      $2.7 \gev$ &
      \phantom{0}$9.7 \pm 0.6$ ~~ ($69 / 57$) &
      $9.2 \pm 0.7$ ~~ ($11 / 12$)
      \\
      \\
      \multicolumn{3}{c}{$\gamma \, p \to K^{+} \Sigma ^{0}$} \\
      $\sqrt{s_{\rm min}}$ & 
      $n$ ($\chi ^{2} / N_{\rm d.o.f.}$), 5 bins & 
      $n$ ($\chi ^{2} / N_{\rm d.o.f.}$), 1 bin
      \\
      \hline
      $2.3 \gev$ &
      $10.5 \pm 0.2$ ~~ ($387 / 257$) & 
      $9.3 \pm 0.2$ ~~ ($34 / 52$) 
      \\
      $2.4 \gev$ &
      \phantom{0}$9.5 \pm 0.2$ ~~ ($191 / 207$) &
      $8.7 \pm 0.2$ ~~ ($18 / 42$)
      \\
      $2.5 \gev$ &
      \phantom{0}$9.2 \pm 0.2$ ~~ ($83 / 157$) &
      $8.8 \pm 0.3$ ~~ ($16 / 32$) 
      \\
      $2.6 \gev$ &
      \phantom{0}$9.2 \pm 0.2$ ~~ ($35 / 107$) &
      $9.1 \pm 0.4$ ~~ ($10 / 22$)
      \\
      $2.7 \gev$ &
      $10.2 \pm 0.3$ ~~ ($10 / 57$) &
      $9.8 \pm 0.4$ ~~ ($3 / 12$)
      \\
    \end{tabular*}
  \end{ruledtabular}
\end{table}

\begin{table}[t]
  \caption{Fitted values of the scaling factor $n$ for $\gamma \, p \to
    K^{+} \Lambda (1405)$, $K^{+} \Sigma (1385)^{0}$, and $K^{+}
    \Lambda (1520)$.  We also show the $\chi ^{2}$ values divided
    by the degrees of freedom $N_{\rm d.o.f.}$.  }
  \label{tab:n2}
  \begin{ruledtabular}
    \begin{tabular*}{8.6cm}{@{\extracolsep{\fill}}lll}
      \multicolumn{3}{c}{$\gamma \, p \to K^{+} \Lambda (1405)$} \\
      $\sqrt{s_{\rm min}}$ & 
      $n$ ($\chi ^{2} / N_{\rm d.o.f.}$), 6 bins & 
      $n$ ($\chi ^{2} / N_{\rm d.o.f.}$), 2 bins
      \\
      \hline
      $2.3 \gev$ &
      $10.7 \pm 0.3$ ~~ ($36 / 29$) &
      $10.2 \pm 0.5$ ~~ ($8 / 9$)
      \\
      $2.4 \gev$ &
      $11.3 \pm 0.4$ ~~ ($28 / 23$) &
      $10.3 \pm 0.7$ ~~ ($7 / 7$) 
      \\
      $2.5 \gev$ &
      $10.6 \pm 0.6$ ~~ ($20 / 17$) &
      \phantom{0}$9.0 \pm 1.1$ ~~ ($4 / 5$) 
      \\
      $2.6 \gev$ &
      \phantom{0}$9.6 \pm 1.0$ ~~ ($11 / 11$) &
      \phantom{0}$7.6 \pm 2.2$ ~~ ($4 / 3$)
      \\
      $2.7 \gev$ &
      \phantom{0}$6.5 \pm 1.7$ ~~ ($3 / 5$) &
      \phantom{0}$5.2 \pm 4.7$ ~~ ($1 / 1$)
      \\
      \\
      \multicolumn{3}{c}{$\gamma \, p \to K^{+} \Sigma (1385)^{0}$} \\
      $\sqrt{s_{\rm min}}$ & 
      $n$ ($\chi ^{2} / N_{\rm d.o.f.}$), 6 bins & 
      $n$ ($\chi ^{2} / N_{\rm d.o.f.}$), 2 bins
      \\
      \hline
      $2.3 \gev$ &
      $10.4 \pm 0.2$ ~~ ($74 / 29$) & 
      $10.5 \pm 0.2$ ~~ ($5 / 9$) 
      \\
      $2.4 \gev$ &
      $10.9 \pm 0.3$ ~~ ($52 / 23$) &
      $10.7 \pm 0.3$ ~~ ($4 / 7$)
      \\
      $2.5 \gev$ &
      $11.4 \pm 0.4$ ~~ ($34 / 17$) &
      $10.2 \pm 0.4$ ~~ ($2 / 5$) 
      \\
      $2.6 \gev$ &
      $12.5 \pm 0.7$ ~~ ($20 / 11$) &
      $10.1 \pm 0.9$ ~~ ($2 / 3$)
      \\
      $2.7 \gev$ &
      $12.2 \pm 1.6$ ~~ ($7 / 5$) &
      \phantom{0}$9.6 \pm 0.3$ ~~ ($0 / 1$)
      \\
      \\
      \multicolumn{3}{c}{$\gamma \, p \to K^{+} \Lambda (1520)$} \\
      $\sqrt{s_{\rm min}}$ & 
      $n$ ($\chi ^{2} / N_{\rm d.o.f.}$), 6 bins & 
      $n$ ($\chi ^{2} / N_{\rm d.o.f.}$), 2 bins
      \\
      \hline
      $2.3 \gev$ &
      $9.0 \pm 0.2$ ~~ ($87 / 29$) & 
      \phantom{0}$9.6 \pm 0.4$ ~~ ($21 / 9$) 
      \\
      $2.4 \gev$ &
      $9.4 \pm 0.3$ ~~ ($57 / 23$) &
      $10.0 \pm 0.7$ ~~ ($18 / 7$)
      \\
      $2.5 \gev$ &
      $9.8 \pm 0.5$ ~~ ($40 / 17$) &
      \phantom{0}$9.3 \pm 1.1$ ~~ ($14 / 5$) 
      \\
      $2.6 \gev$ &
      $8.7 \pm 0.7$ ~~ ($22 / 11$) &
      \phantom{0}$6.6 \pm 0.8$ ~~ ($2 / 3$)
      \\
      $2.7 \gev$ &
      $7.7 \pm 2.1$ ~~ ($18 / 5$) &
      \phantom{0}$5.8 \pm 1.9$ ~~ ($1 / 1$)
      \\
    \end{tabular*}
  \end{ruledtabular}
\end{table}

In this study, we analyze the photoproductions of $\Lambda (1405)$,
$\Sigma (1385)^{0}$, and $\Lambda (1520)$.  Especially, $\Lambda
(1405)$ is speculated to be a $\bar{K} N$ bound state or possibly a
mixture of $\bar{K} N$ bound state, compact five-quark system, and
conventional three-quark one, so that its number of constituents might
be different from that for an ordinary baryon ($n_{\Lambda (1405)} \ne
n_\Lambda=3$).  The results are shown in Figs.~\ref{fig:KpL1405},
\ref{fig:KpS1385}, and \ref{fig:KpL1520}. As obvious from these
figures, there are not enough data, and the maximum energy is
$\sqrt{s}=$2.8 GeV in comparison with 3.5 GeV in Figs.~\ref{fig:KpL}
and \ref{fig:KpS}.  It is apparent from Figs.~\ref{fig:KpL1405},
\ref{fig:KpS1385}, and \ref{fig:KpL1520} that the data are scarce and
the scaling property is less clear.  Nevertheless, we tried to fit the
cross section at high energies by the scaling behavior.

Fitted factors $n$ are shown in Table~\ref{tab:n2} for six-bin data as
well as two-bin data and with various $\sqrt{s_{\rm min}}$ values from
2.3 GeV to 2.7 GeV.  The results are shown in Fig.~\ref{fig:n_Ystar}
by taking $\sqrt{s_{\rm min}}$ as the abscissa. The factors $n$ have
large errors especially if the data are limited in the high-energy
region, namely $\sqrt{s} \ge \sqrt{s_{\rm min}}=2.5$, 2.6, or 2.7 GeV.
For example, the factors $n$ for $\Lambda (1405)$ in
Fig.~\ref{fig:n_Ystar} indicate that $\Lambda (1405)$ looks like a
five-quark state ($n=11$) if the data are used from $\sqrt{s_{\rm
    min}}=2.3$ or 2.4 GeV, whereas the factor becomes smaller $n<11$
with $\sqrt{s_{\rm min}} \ge 2.5 \gev$.  For $\Sigma (1385)^{0}$, the
scaling factor is close to $n=11$ if the data with $\sqrt{s_{\rm
    min}}=2.3$--$2.5$ GeV are used; however, it becomes uncertain if
only the data at higher energies, $\sqrt{s_{\rm min}}=2.6$ and $2.7
\gev$, are analyzed. In terms of scaling behavior at $\sqrt{s_{\rm
    min}} \ge 2.5$ GeV, $\Lambda(1405)$ and $\Lambda(1520)$ seem to be
more consistent with three-quark system while $\Sigma (1385)^{0}$ is
close to five-quark one.  Suppose a baryon is a mixture of
conventional three-quark system, compact five-quark one, and diffuse
meson--baryon molecule.  Then, a transition in the scaling behavior at
different energy region might be expected and the production at very
high energies could be dominated by the state of minimal $n$, i.e.,
the relatively tight three-quark component ($n=9$). Such a scenario
seems favored by the observed $\sqrt{s_{\rm min}}$ dependence of
scaling behavior for $\Lambda(1405)$.  However, conclusive information
of the scaling factors $n$ for the photoproductions of $\Lambda
(1405)$, $\Sigma (1385)^{0}$, and $\Lambda (1520)$ cannot be obtained
from the current data, as seen from the scattered data even at high
energies in Figs.~\ref{fig:KpL1405}, \ref{fig:KpS1385}, and
\ref{fig:KpL1520}.

From our analysis, we propose that experimentalists measure the
high-energy region ($\sqrt{s} > 2.8$ GeV) of the photoproduction cross
sections to draw a solid conclusion for the scaling factor $n$.  It
should be especially important for an exotic hadron candidate,
$\Lambda (1405)$. This kind of experiments should be possible at JLab
after the 12-GeV upgrade of its accelerator \cite{jlab-12gev}.
Similar measurements could be also done at LEPS (Laser Electron Photon
beamline at SPring-8) II, especially because there is a plan to have a
detector for large-angle scattering measurements \cite{LEPS-future}
with the increase of photon energy.  These high-energy experiments are
important as a new direction of exotic hadron studies, which have been
done so far in low-energy global observables such as spins, parities,
masses, and decay widths.

\section{Summary}
\label{summary}

We analyzed the exclusive photoproduction data of hyperons, 
$\gamma \, p \to K^+ Y$ [$Y=\Lambda$, $\Sigma^0$, $\Lambda (1405)$, $\Sigma
  (1385)^{0}$, or $\Lambda (1520)$] by using the constituent-counting
rule at high energies for determining the number of constituents involved
in the exclusive reactions. Our studies intend to clarify
the internal constituents of an exotic hadron candidate $\Lambda
(1405)$, because the counting rule is considered to be valid for
hadrons and nuclei, such as the proton, the pion, and
deuteron. Therefore, our work can shed light on a new method of
exotic-hadron studies for clarifying their internal structure by
high-energy reactions where quark and gluon degrees of freedom are
valid.

Our analysis results indicate that the data for the reactions $\gamma
\, p \to K^+ \Lambda$ and $\gamma \, p \to K^+ \Sigma^0$ are
consistent with the factor $n=9$. It suggests that $\Lambda$ and
$\Sigma^0$ have three quark constituents in them, which is consistent
with conventional quark models. On the other hand, at this moment it
is difficult to pin down the factor $n$ for the photoproductions of
$\Lambda (1405)$, $\Sigma (1385)^{0}$, and $\Lambda (1520)$ due to
lack of data.  Nevertheless, the observed scaling behavior of hyperon
$\Lambda (1405)$ seems to suggest that it is not purely three-quark or
five-quark system.  The number of constituent quarks for the excited
hyperons could be determined by their photoproduction processes in
principle if enough data are taken in a wide energy region.  If an
excited hyperon is a mixture of three-quark and five-quark states,
energy dependence of the scaling behavior could provide valuable
information for its composition and mixture.

The current method was not applied to exotic hadrons until recently,
and it should be a new approach for identifying exotic hadrons and for
clarifying their internal structure.  The scaling for exotic hadron
candidates should be clarified by precise cross-section measurements
in future for the photoproductions in the high-energy region
($\sqrt{s} > 2.8$ GeV) and also hadronic exclusive reactions, for
example, at J-PARC (Japan Proton Accelerator Research Complex)
\cite{j-parc-exp}.

\section*{Acknowledgments}

This work is partly supported by Grants-in-Aid for Scientific Research
from MEXT and JSPS (No.~15K17649, No.~15J06538, and No.~25105010).



\end{document}